\documentclass[aps,amsmath,amssymb,twocolumn,10pt,superscriptaddress,pre,longbibliography]{revtex4-1}
\usepackage{graphicx}  
\usepackage{bbold}
\usepackage[dvipsnames]{xcolor}
\usepackage[colorlinks=true,linkcolor=blue,citecolor=red,urlcolor=magenta]{hyperref}
\usepackage{orcidlink} 
\usepackage{charter}
 
\begin{document}

\title{Quantum walker trapped by self-similarity of the Sierpi\'nski carpet}

\author{Tomasz Sowi\'nski\;\orcidlink{0000-0002-7970-4371}}
\affiliation{Institute of Physics, Polish Academy of Sciences, Aleja Lotnikow 32/46, PL-02668 Warsaw, Poland}

\begin{abstract}
We study the dynamics of a single quantum particle on a finite-size square lattice with a fractal structure resembling the Sierpi\'nski carpet, and compare it to the dynamics on a uniform lattice of the same size. For a particle initially localized at a corner of the lattice, we monitor the probability of finding it near the initial and opposite corners using zone-integrated probabilities, allowing a consistent comparison across fractal orders. While on the uniform lattice the particle reaches the opposite corner ballistically, in a time proportional to the lattice size, on the Sierpi\'nski lattice it becomes increasingly confined to the vicinity of its initial position. We show that this trapping builds up self-similarly across the whole hierarchy of corner zones of the lattice.
\end{abstract} 
\maketitle 
 
\section{Introduction}

Quantum walkers spread through a discrete lattice in a fundamentally different way than their classical counterparts. Quantum interference between different paths typically leads to ballistic rather than diffusive transport, and this speed-up underlies their prominent role in quantum algorithms and in modeling coherent energy and information transfer~\cite{Aharonov1993,Farhi1998,2003Ambainis,Kempe2003}. 

On theoretical footing, quantum walks are considered in two distinct, complementary frameworks. In the discrete-time quantum walk \cite{Aharonov1993,Childs2004,Inui2004,Inui2005,2004Wojcik,2010Kitagawa,Wojcik2012,2013Cedzich,2015PhysRevA,2018Perez,2020ArnaultPRA,2024Wojcik}, a walker evolves stroboscopically, through repeated application of a unitary coin operator, acting on an internal, auxiliary degree of freedom, followed by a shift operator that moves the particle conditioned on the coin state. Continuous-time quantum walk, in contrast, requires no auxiliary coin. The dynamics is governed directly by the Schr\"odinger equation with a time-independent, tight-binding Hamiltonian, exactly as for a single particle hopping between the sites of a lattice~\cite{Farhi1998,2010WitthautPRA,Mulken2011,Perets2008,Agliari2010,Jaczewski2024}. The two approaches, although connected~\cite{2006Strauch}, are not simply different numerical schemes for the same physics, but they can exhibit qualitatively different transport and localization properties on the same underlying graph. 

How this picture changes when the underlying lattice is not regular but has a fractal, self-similar structure is a natural and still actively studied question. Given that such fractal structures are now realizable in a variety of experimental platforms, including photonic waveguide arrays~\cite{Xu2021,Biesenthal2022}, scanning-tunneling-microscopy-assembled electronic lattices~\cite{Kempkes2019,Shang2015}, and cold atoms trapped in optical tweezer arrays~\cite{Tian2023}, these kinds of problems have become interesting not only from a theoretical perspective, but also in terms of practical applications.

In the group of two-dimensional fractals, the Sierpi\'nski ones have become the paradigmatic testbeds for this kind of question. For the Sierpi\'nski gasket, both discrete- \cite{Patel2012,Lara2018,Tamegai2018,Sato2020,Mares2020,ZhaoYang2022} as well as continuous-time \cite{Domany1983,Wang1995,Darazs2014,Dhamapurkar2026} quantum walks have been studied extensively and many interesting dynamical features were identified and explored. In contrast, the Sierpi\'nski carpet, the two-dimensional fractal we focus on in this work, has received comparatively less attention, and the emerging picture is considerably less settled. Depending on the specific study, the reported transport behavior ranges from a trend towards trapping and suppressed conductance in extended energy windows~\cite{Darazs2014,vanVeen2016} to pronounced, super-diffusive speed-up relative to the classical random walk~\cite{Xu2021,RojoFrancas2024}. Part of this disagreement has been traced to seemingly minor differences in how the lattice is constructed and, in particular, in which pairs of sites are connected by a tunneling matrix element~\cite{RojoFrancas2024}. Related, geometry-induced, disorder-free trapping has also very recently been reported for electrons on other fractal lattices~\cite{Pal2025}.

In this work we take a closer look at the phenomenon of particle trapping induced by a fractal structure of the Sierpi\'nski carpet, within continuous-time dynamics. We analyze how quickly a particle, initially localized at a corner of the lattice, is able to leave the region it initially occupies, as a function of the size of the whole system. As a reference system we use a uniform lattice of the same edge length, and we introduce coarse-grained quantities that better capture the transport properties. We further generalize this approach to a whole hierarchy of self-similar corner zones of increasing size, which lets us track how the trapping builds up across different length scales of the fractal. This reveals that, as the fractal structure grows, ballistic transport breaks down and the particle becomes trapped near its initial position. 

The paper is organized as follows. In Sec.~\ref{Section2} we introduce the Sierpi\'nski lattice and explain its construction. In Sec.~\ref{Section3} we define the Hamiltonian governing the dynamics and the observables used to characterize the transport. Section~\ref{Section4} presents our numerical results for the trapping of the particle. In Sec.~\ref{Section5} we extend this analysis to the whole hierarchy of self-similar zones, and we summarize our findings and discuss their implications in Sec.~\ref{Section6}.

\section{The Sierpi\'nski lattice} \label{Section2}
We focus on finite-size, square-shaped lattices resembling the Sierpi\'nski carpet~\cite{sierpinski1916}. The lattice is completely defined by giving a single integer ${\cal R}$, which we call the fractal order. The construction is as follows. For a given ${\cal R}$, we consider a uniform square lattice of $9^{\cal R}$ sites (the edge size of the lattice is $D=3^{\cal R}$). Then the lattice is divided into $9$ equal squares of size $D/3$ and all sites belonging to the middle square are removed. The procedure is repeated for all remaining squares until there are no more sites to be removed. The resulting lattice contains $N=8^{\cal R}$ sites and in the limit ${\cal R}\rightarrow \infty$ reproduces the Sierpi\'nski carpet fractal. In Fig.~\ref{Fig1} we visualise Sierpi\'nski lattices defined in this way for several fractal orders  ${\cal R}$. By definition, lattice sites are enumerated by pairs of positive integers $(\mathbb{x},\mathbb{y})$ indicating their coordinates on a plane, $1\leq \mathbb{x},\mathbb{y} \leq 3^{\cal R}$. For further convenience, we will also use an algebraic, two-integer vector $\vec{\mathbb{r}}=(\mathbb{x},\mathbb{y})$ representing these coordinates.

\section{The dynamics} \label{Section3}
We consider the continuous-time dynamics of a single quantum particle moving on the Sierpi\'nski lattice and we compare it with the dynamics on uniform lattice. The corresponding Hilbert spaces are spanned by quantum states $\{|\vec{\mathbb{r}}\rangle\}$ describing particle occupying lattice site labelled by vector $\vec{\mathbb{r}}=(\mathbb{x},\mathbb{y})$. Of course, the quantum state of a particle at any moment $|\psi(t)\rangle$ can be decomposed in this basis as
\begin{equation}
|\psi(t)\rangle = \sum_{\vec{\mathbb{r}}} \psi(\vec{\mathbb{r}},t) |\vec{\mathbb{r}}\rangle,
\end{equation}
where $\psi(\vec{\mathbb{r}},t)$ has a natural interpretation of the probability amplitude of finding a particle at time $t$ in the lattice site $\vec{\mathbb{r}}$. In the following, we assume the simplest scenario in which the particle can tunnel only to a neighbouring lattice site, {\it i.e.}, when particle occupies site $(\mathbb{x},\mathbb{y})$ it can tunnel only to one of sites $(\mathbb{x}\pm 1,\mathbb{y})$ or $(\mathbb{x},\mathbb{y}\pm 1)$ provided that this particular site belongs to the lattice (was not removed during the construction). It simply means that the dynamics is governed by the lattice Hamiltonian of the form
\begin{equation}
{\cal H} = -J \sum_{\{\vec{\mathbb{r}},\vec{\mathbb{r}}'\}}|\vec{\mathbb{r}}'\rangle\langle \vec{\mathbb{r}}|,
\end{equation}
where summation runs only over neighbouring sites as defined above, and $J$ is the tunneling amplitude, which also defines the energy unit in the problem.
\begin{figure}
\includegraphics[width=\linewidth]{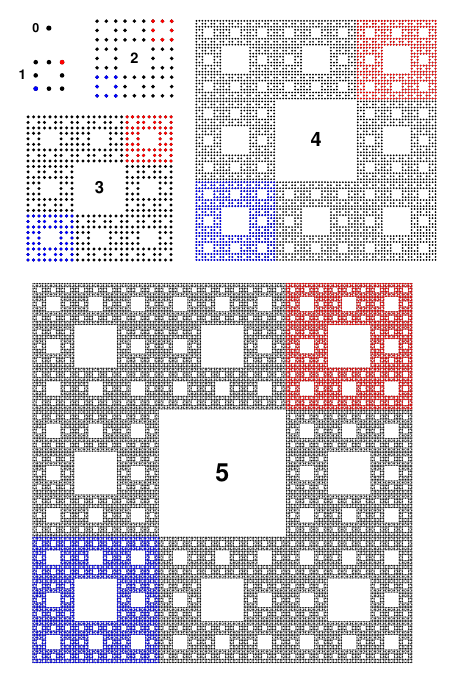}
\caption{Visualization of the Sierpi\'nski lattices for different fractal orders ${\cal R}$. Number of lattice sites as well as the edge size (number of sites along the edge) exhibit power-law scaling with the fractal order, $N=8^{\cal R}$ and $D=3^{\cal R}$, respectively. The initial zone $\mathtt{Z_{INI}}$ and the final zone $\mathtt{Z_{FIN}}$ are marked with blue and red color, respectively.
\label{Fig1}}
\end{figure}

\begin{figure}
\includegraphics[width=\linewidth]{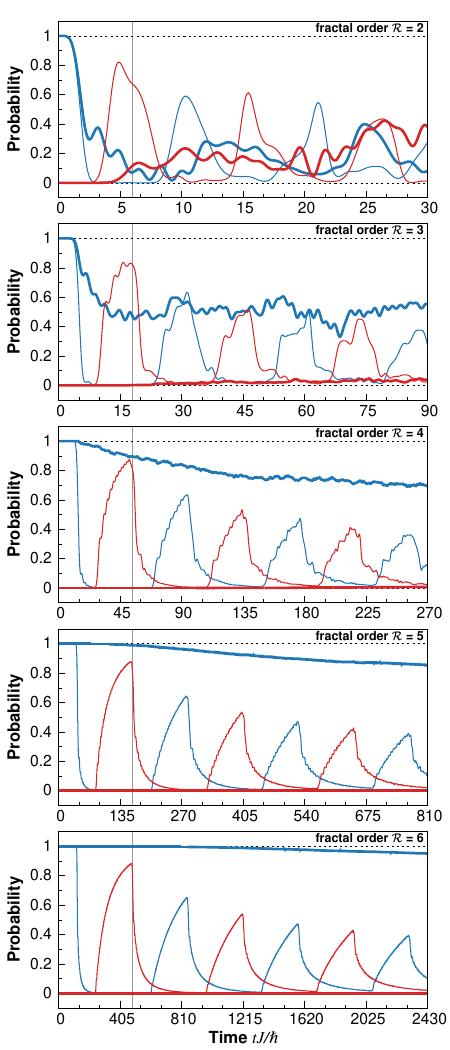}
\caption{Probability of finding the particle in the initial zone $\mathtt{Z_{INI}}$ (blue) and the final zone $\mathtt{Z_{FIN}}$ (red) zone of the lattice in the case of Sierpi\'nski (thick) and uniform (thin) lattice. In all the cases particle is initially localised in the corner of the lattice belonging to the initial zone. Clearly, for increasing fractal order ${\cal R}$,  the particle finds it increasingly difficult to reach the opposite zone of the lattice and remains trapped in the initial zone. Contrarily, for the uniform lattice, the particle reaches the opposite zone in time proportional to the edge size of the lattice, $T_U \approx 2\hbar D/3J$ marked with a vertical grey line. 
\label{Fig2}}
\end{figure}

We aim to study the motion of the quantum particle, which initially is perfectly localized in the corner site (without losing generality, we choose the bottom left) of the lattice, $\vec{\mathbb{r}}_0=(1,1)$, and compare its quantum diffusion on the Sierpi\'nski lattice and the uniform lattice. The dynamics is governed by the Schr\"odinger equation 
\begin{equation}
i\hbar\frac{\mathrm{d}}{\mathrm{d}t}|\psi(t)\rangle = {\cal H}|\psi(t)\rangle
\end{equation}
which we solve numerically with the fourth-order Runge-Kutta method.

Several quantities make the comparison of the dynamics on different lattices clear and quantitative. Firstly, the simplest approach is to study the time evolution of the survival probability ${\cal S}_0(t) = |\psi(\vec{\mathbb{r}}_0,t)|^2$, {\it i.e.}, the probability that the particle remains occupying the initial site. Similarly, one can monitor the probability of finding the particle on the opposite corner of the lattice ${\cal F}_0(t) = |\psi(\vec{\mathbb{r}}_F,t)|^2$, where $\vec{\mathbb{r}}_F=(D,D)$. In this way, one can estimate the time needed to reach the opposite side of the lattice. Both quantities, although uniquely defined, are not appropriate when we want to compare dynamics on lattices of different sizes, especially when the lattice has no trivial internal fractal-like structure. It is much better to compare conceptually similar, but mesoscopic features that scale appropriately with the lattice size. Due to the natural structure of the Sierpi\'nski lattice, we identify two zones on opposite sides of the lattice: the initial zone $\mathtt{Z_{INI}}$ localized in the corner of the lattice containing the initial site $\vec{\mathbb{r}}_0$ and the final zone $\mathtt{Z_{FIN}}$ localized on the opposite corner. Both zones are marked in Fig.~\ref{Fig1} with blue and red colors, respectively. Both zones are chosen as one of the nine equal squares of size $D/3$ appearing at the first level of the lattice construction, {\it i.e.}, each zone always contains a fixed fraction ($1/8$ and $1/9$ respectively for Sierpi\'nski and uniform lattice) of all lattice sites, irrespective of the fractal order ${\cal R}$. Correspondingly, we define the zone-integrated probabilities
\begin{subequations}
\begin{align}
{\cal S}(t) &= \sum_{\vec{\mathbb{r}}\in \mathtt{Z_{INI}}} |\langle\vec{\mathbb{r}}|\psi(t)\rangle|^2, \\
{\cal F}(t) &= \sum_{\vec{\mathbb{r}}\in \mathtt{Z_{FIN}}} |\langle\vec{\mathbb{r}}|\psi(t)\rangle|^2.
\end{align}
\end{subequations}
Unlike their single-site counterparts ${\cal S}_0(t)$ and ${\cal F}_0(t)$, these zone-integrated probabilities scale consistently with the lattice size and fractal order ${\cal R}$.

\section{Trapping in the fractal lattice} \label{Section4}

In Fig.~\ref{Fig2} we present the time evolution of the probabilities ${\cal S}(t)$ (thick and thin blue lines) and ${\cal F}(t)$ (thick and thin red lines) obtained for the Sierpi\'nski lattice and the uniform lattice, respectively, for several fractal orders ${\cal R}=2,\ldots,6$. For the uniform lattice, the dynamics has a simple, essentially ballistic character. The particle leaves the initial site almost immediately and the probability ${\cal F}(t)$ rises to a pronounced maximum close to unity at a time consistent with $T_U\approx 2\hbar D/3J$ (vertical grey line), {\it i.e.}, scaling linearly with the lattice size $D$. At later times both probabilities exhibit a sequence of damped, quasi-periodic revivals, a signature of the wave packet reflecting off the boundaries of the finite lattice. To demonstrate that there is complete regularity here and that the pattern scales with the size of the system, in Fig.~\ref{Fig2} we use appropriately scaled time ranges on the time axis (the time range is multiplied by factor 3 for consecutive orders).

The dynamics on the Sierpi\'nski lattice is qualitatively different, and the difference becomes more pronounced with increasing fractal order. At sufficiently long times and for the smallest order shown, ${\cal R}=2$, the two zones become comparably populated and the fractal structure only mildly perturbs the transport. Already for ${\cal R}=3$, however, the behavior changes markedly. The integrated survival probability ${\cal S}(t)$ remains close to unity and decays only slowly and monotonically, without any sharp initial drop, while ${\cal F}(t)$ stays close to zero for the entire simulated time. It means that the particle is not able to reach the final zone. This asymmetry grows systematically and rapidly with ${\cal R}$, and for ${\cal R}=5$ and ${\cal R}=6$ the particle never leaves the initial zone within the accessible simulation time, {\it i.e.}, ${\cal S}(t)$ remains essentially pinned near unity. In other words, increasing the fractal order does not simply slow down the transport but it progressively suppresses it, effectively trapping the particle in the vicinity of its initial position.

This behavior can be understood phenomenologically as a consequence of the self-similar, hierarchical geometry of the Sierpi\'nski lattice. By construction, as the fractal order ${\cal R}$ increases, there is a substantial rise of paths connecting the initial and final corners requiring multiple passes through the narrow necks of the lattice, {\it i.e.}, single rows or columns of sites bordering the removed squares. Furthermore, this occurs at every length scale present in the fractal, from the largest removed square down to the smallest. Each such neck acts as an effective partial barrier for the propagating wave packet, and its presence at every hierarchical level means that the number of such barriers a particle needs to traverse grows with the fractal order ${\cal R}$. Treating the consecutive necks as a series of (weakly) reflective barriers, the overall transmission is expected to decrease roughly exponentially with the number of barriers, and hence with ${\cal R}$. This provides a simple qualitative explanation for why the crossing probability ${\cal F}(t)$ becomes vanishingly small already for moderate fractal orders, even though the lattice remains fully connected and no site-to-site links are actually removed from the accessible paths. From this perspective, this effect is purely deterministic and geometric in origin.

\section{Scale dependence of the trapping} \label{Section5}

Having established that the particle becomes trapped in the corner zones of fixed relative size, it is natural to ask how this trapping depends on the size of the zone itself, relative to the size of the whole lattice. Therefore, now we generalize the definition of the initial zone introduced above into a whole hierarchy of self-similar initial zones $\mathtt{Z_{INI}^{\cal K}}$, obtained by the same fractal construction restricted to a sub-lattice of edge size $3^{\cal K}$ containing the initial site ($0\leq{\cal K}\leq{\cal R}$). The corresponding zone-integrated probabilities are denoted as ${\cal S}_{\cal K}(t)$. Of course, two extreme cases ${\cal K}=0$ and ${\cal K}={\cal R}-1$ reproduce probabilities ${\cal S}_0(t)$ and ${\cal S}(t)$ defined in Sec.~\ref{Section3}. 

\begin{figure}
\includegraphics[width=\linewidth]{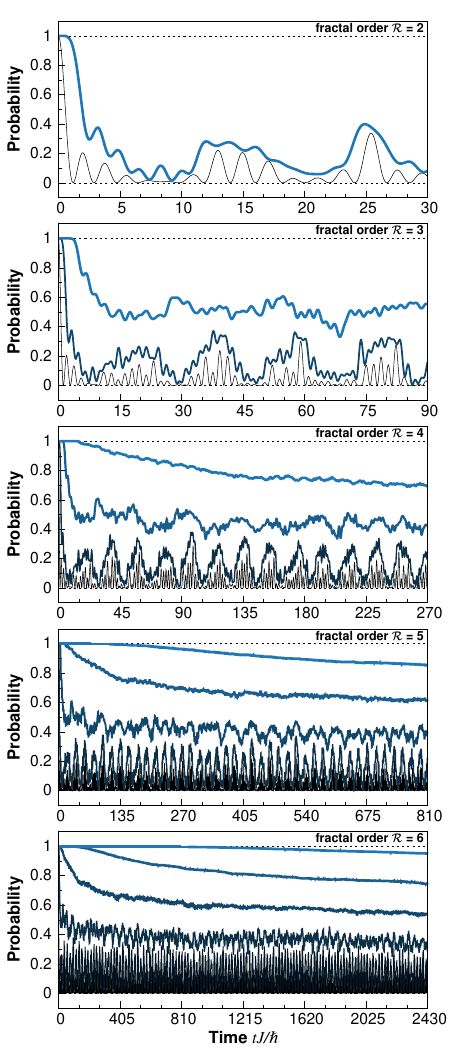}
\caption{Time evolution of the initial-zone probabilities ${\cal S}_{\cal K}(t)$ on the Sierpi\'nski lattice, for the hierarchy of corner zones of increasing size $3^{\cal K}$, ${\cal K}=0,1,\ldots,{\cal R}-1$ (darker to lighter shades of blue), for the same fractal orders ${\cal R}=2,\ldots,6$ as in Fig.~\ref{Fig2}. The lightest curve in each panel, corresponding to ${\cal K}={\cal R}-1$, coincides with the thick blue curve of Fig.~\ref{Fig2}.
\label{Fig3}}
\end{figure}
The results shown in Fig.~\ref{Fig3} reveal a clear and systematic trend. The smallest zones, ${\cal K}=0$ and ${\cal K}=1$, are quickly and repeatedly emptied and refilled, {\it i.e.}, the corresponding probabilities drop close to zero on a short time scale and subsequently oscillate with a comparatively large amplitude, indicating that the particle only transiently visits the closest neighborhood of the initial site before spreading further into the surrounding structure. As ${\cal K}$ increases, the zones retain the probability increasingly effectively: the initial drop becomes less pronounced, the oscillations get damped, and the long-time value of ${\cal S}_{\cal K}(t)$ systematically increases, culminating in the largest zone, ${\cal K}={\cal R}-1$, which -- consistently with Fig.~\ref{Fig2} -- retains almost all of the probability for large ${\cal R}$. In other words, trapping is not an all-or-nothing effect tied to a particular length scale, but builds up progressively and self-similarly across the whole hierarchy of zones. It means that the particle is not simply confined to a region of some fixed size, but is instead reluctant to leave any of the self-similar neighborhoods of its initial position, with the reluctance growing systematically with the size of the neighborhood considered. This is fully consistent with, and lends further support to, the bottleneck picture proposed above. Escaping the initial zone of size $3^{\cal K}$ requires crossing bottlenecks associated with hierarchical levels up to and including ${\cal K}$, so that a larger zone is, by construction, protected by a larger number of such barriers.

To make this analysis more complete, additionally we introduce the characteristic escape time $T_{\cal K}$, defined as the first instant at which the probability of finding the particle outside the ${\cal K}$-order initial zone $\mathtt{Z_{INI}^{\cal K}}$ exceeds a fixed threshold (in our work we set this threshold on $3\%$). We expect the dependence of this characteristic time on ${\cal K}$ to be markedly different for the fractal and the uniform lattice, since only the former is affected by the trapping mechanism discussed above. It is natural to rescale $T_{\cal K}$ by the linear size of the zone, $3^{\cal K}$, since this is the length scale the particle actually has to traverse to escape.

\begin{figure}
\includegraphics[width=\linewidth]{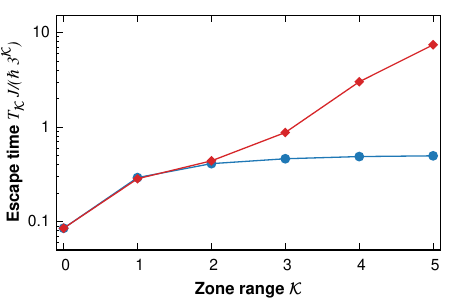}
\caption{Rescaled escape time $T_{\cal K}J/(\hbar\, 3^{\cal K})$ as a function of zone range ${\cal K}$, for the uniform (blue) and the Sierpi\'nski (red) lattice of fractal order ${\cal R}=6$. Note the significant difference in their dependence on zone range ${\cal K}$.
\label{Fig4}}
\end{figure}

In Fig.~\ref{Fig4} we compare the rescaled escape time $T_{\cal K}J/(\hbar\, 3^{\cal K})$, as a function of ${\cal K}$, obtained for the largest lattice considered, ${\cal R}=6$, on both the Sierpi\'nski and the uniform lattice. The two curves shown start from essentially the same value at ${\cal K}=0$, as expected, since at this level the initial zone reduces to a single site and the two lattices are locally indistinguishable. Along with increasing ${\cal K}$, the rescaled escape time quickly saturates to an essentially constant value when the dynamics on a uniform lattice is considered. This is exactly what one expects for ballistic transport, $T_{\cal K}\propto 3^{\cal K}$. The flat blue curve in Fig.~\ref{Fig4} is thus a direct confirmation, on a quantitative level, that transport on the uniform lattice does not know about any built-in length scale other than the ballistic velocity itself.

The dynamics on the Sierpi\'nski lattice behaves in a qualitatively different way. Even after the same rescaling by $3^{\cal K}$, the escape time keeps growing with ${\cal K}$, roughly exponentially, over the whole accessible range. This means that the escape time on the fractal lattice does not simply scale with the zone size, as it does for the uniform lattice, but grows systematically faster. In consequence, the particle takes progressively longer, relative to the size of the zone itself, to escape from larger zones. This is precisely the behavior expected from the hierarchical bottleneck picture proposed already. Escaping a zone of size $3^{\cal K}$ requires crossing an additional barrier at every one of the ${\cal K}$ hierarchical levels contained within it, so the residual, size-independent growth of $T_{\cal K}J/(\hbar\,3^{\cal K})$ with ${\cal K}$ can be understood as the accumulated cost of overcoming these barriers.

\section{Conclusions} \label{Section6}

We studied the dynamics of a single quantum particle initially localized at a corner of a finite, square-shaped fractal lattice resembling the Sierpi\'nski carpet, and compared it to the dynamics on a uniform lattice of the same size. To make this comparison meaningful across lattices of different fractal order, we introduced zone-integrated probabilities quantifying the presence of the particle in the vicinity of the initial and the opposite corner of the lattice.

On the uniform lattice, the dynamics is essentially ballistic and the particle reaches the opposite corner of the lattice at a well-defined time $T_U\approx 2\hbar D/3J$, scaling linearly with the lattice size $D$. The subsequent evolution shows damped revivals due to reflections from the lattice boundaries. The dynamics on the Sierpi\'nski lattice is markedly different, and the difference becomes more prominent for larger fractal orders ${\cal R}$. In this case, the particle becomes increasingly confined to the initial zone, while the probability of reaching the opposite corner is progressively suppressed. Already for moderate fractal orders the particle remains, for all practical purposes, trapped near its initial position throughout the entire simulated evolution. Notably, this trapping is not simply a slowed-down version of the ballistic transport observed on the uniform lattice where the characteristic crossing time $T_U$ is a well-defined quantity and scales clearly with the lattice size. On the Sierpi\'nski lattice no such characteristic crossing time emerges. Instead of a rescaled but otherwise similar transport process, the dynamics qualitatively changes character, and for sufficiently large ${\cal R}$ the notion of a crossing time effectively ceases to apply. This absence of a simple time scaling, rather than a mere slowdown, is one of the most interesting features of the transport on the Sierpi\'nski lattice.

We argued that this behavior can be partially interpreted phenomenologically as a consequence of the self-similar, hierarchical geometry of the lattice, in which any path between opposite corners must repeatedly pass through narrow necks present at every length scale of the fractal. Treating these necks as a series of effective partial barriers suggests an overall transmission that decreases roughly exponentially with the fractal order, providing a simple qualitative explanation of the observed trapping. 

The results presented here suggest that lattice geometry alone, in the absence of disorder but in the presence of a fractal structure, can lead to pronounced suppression of quantum transport. Recent advances in the engineering of artificial lattices in photonic and cold-atom platforms provide promising opportunities to investigate the trapping mechanism reported here experimentally. This direction appears particularly promising for future studies, especially in the context of extending the analysis to systems with several interacting quantum walkers, where the interplay between fractal geometry, quantum interference, and interparticle interactions may reveal even richer transport phenomena known for uniform lattice systems~\cite{science1260364,PhysRevA.96.043629,PhysRevA.101.052341,PhysRevA.102.043326,SciRep22056,PhysRevLett.127.100406,PhysRevA.104.033306,PhysRevLett.129.050601,PhysRevA.109.043308,PhysRevB.110.014302}.

\section*{Data availability statementx}
All numerical data presented in this paper are available online~\cite{zenodo}.

\section*{Acknowledgments}
The author thanks Vikash Mittal, Armando P\'erez, and Miguel \'Angel Garc\'ia-March for very fruitful and valuable discussions, and the Instituto Universitario de Matem\'atica Pura y Aplicada (IUMPA), Universitat Polit\`ecnica de Val\`encia, for hospitality. This research was partially supported by the National Science Centre (NCN, Poland) within the OPUS project No.~2023/49/B/ST2/03744. For the purpose of Open Access, the author has applied a CC-BY public copyright licence to any Author Accepted Manuscript version arising from this submission.

\bibliography{biblio}

\end{document}